\documentclass[preprint,superscriptaddress,longbibliography]{revtex4-2}

\usepackage{graphicx}
\usepackage{color}
\usepackage{amsfonts}

\usepackage{isomath}
\usepackage{amsmath}
\usepackage{amsthm}
\usepackage{amsfonts}
\usepackage{amssymb}
\usepackage{fdsymbol}
\usepackage{braket}
\usepackage{gensymb}

\let\oldAA\AA
\renewcommand{\AA}{\text{\normalfont\oldAA}}

\begin{document}
\title{Electric Field-Induced Hole- and Electron-Type Flat Bands in Twisted Double Bilayer Graphene}
\author{Zhihao Jiang}
\affiliation{Department of Physics and Astronomy,  Aarhus University, 8000 Aarhus C, Denmark}
\author{Dongkyu Lee}
\affiliation{Department of Physics, University of Seoul, Seoul 02504, Korea}
\author{Alfred J. H. Jones}
\affiliation{Department of Physics and Astronomy,  Aarhus University, 8000 Aarhus C, Denmark}
\author{Youngju Park}
\affiliation{Department of Physics, University of Seoul, Seoul 02504, Korea}
\author{Paulina Majchrzak}
\author{Kimberly Hsieh}
\author{Chakradhar Sahoo}
\author{Thomas S. Nielsen}
\affiliation{Department of Physics and Astronomy,  Aarhus University, 8000 Aarhus C, Denmark}
\author{Kenji Watanabe}
\affiliation{Research Center for Electronic and Optical Materials, National Institute for Materials Science, 1-1 Namiki, Tsukuba 305-0044, Japan}
\author{Takashi Taniguchi}
\affiliation{Research Center for Materials Nanoarchitectonics, National Institute for Materials Science,  1-1 Namiki, Tsukuba 305-0044, Japan}
\author{Philip Hofmann}
\affiliation{Department of Physics and Astronomy,  Aarhus University, 8000 Aarhus C, Denmark}
\author{Yong P. Chen}
\affiliation{Department of Physics and Astronomy,  Aarhus University, 8000 Aarhus C, Denmark}
\author{Jeil Jung}
\affiliation{Department of Physics, University of Seoul, Seoul 02504, Korea}
\author{S{\o}ren Ulstrup}
\email{ulstrup@phys.au.dk}
\affiliation{Department of Physics and Astronomy,  Aarhus University, 8000 Aarhus C, Denmark}

\pagebreak

\begin{abstract}
Application of a finite electric field and small twist angle between two graphene bilayers leads to widely tunable electronic structures that support emergent phenomena.  Here,  we measure the electronic structure of twisted double bilayer graphene with twist angles of 3.1$^{\circ}$ and 6.0$^{\circ}$ using micro-focused angle-resolved photoemission spectroscopy (microARPES) on \textit{in situ} gated samples.  At a twist angle of 6.0$^{\circ}$,  we find that the dispersion around the valence and conduction band extrema is unchanged compared to separate graphene bilayers,  whereas electric field tunable hybridisation gaps and band flattening effects occur at higher binding energies.  In the 3.1$^{\circ}$ structure,  we find that the resulting superlattice drives dispersion changes around the valence and conduction band edges.  Increasing the gate-induced doping in the bottom bilayer leads to the formation of flat bands between the valence band maximum in the bottom bilayer and conduction band minimum in the top bilayer.  Our experiments establish combinations of twist angle and electric field that potentially support strong electron-hole excitations and that provide a benchmark for the theoretical description of moderately twisted double bilayer graphene.
\\ 
\end{abstract}

\maketitle

\section{Introduction}
The realization of precision-engineered moiré superlattices by twisting and stacking two-dimensional (2D) materials has opened a new paradigm in designing electronic correlations,  superconductivity and topological phase transitions, as well as non-linear transport and optical phenomena \cite{Kennes:2021,Andrei:2021,Feng:2021,Luojun:2024}.  Twisted bilayer graphene with the magic twist angle of  1.1$^{\circ}$ constitutes a highly tunable platform that elevates our understanding of electronic structure and transport phenomena underpinning competing correlated insulator and superconducting states in a manner that is very similar to the high-temperature superconductors \cite{Cao2018,Cao:2018b,Andrei:2020}.  Interestingly,  twisted double bilayer graphene (tDBG),  which is composed of two Bernal-stacked (AB) graphene bilayers with a twist angle between the AB layers,  also develops a series of controllable insulating,  magnetic and superconducting states \cite{Shen:2020,Liu2020,He2021,Su2023}.  This appears rather surprising,  as AB bilayer graphene exhibits hyperbolic bands separated by a band gap when subjected to an out-of-plane electric field,  thereby representing a significant departure from the situation in monolayer graphene, which exhibits gapless linear Dirac bands \cite{Neto:2009}.  However,  by applying a displacement field across the AB layers in tDBG with twist angles around the magic angle,  moiré hybridisation effects induce flat bands that are separated from the main conduction and valence band manifolds via hybridisation gaps \cite{Burg:2019,Lee2019,Chebrolu:2019,Nunn:2023,Zhihao:2025,sasmal:2025},  which essentially leads to the same low-energy electron transport phenomena as seen in twisted bilayer graphene \cite{Shen:2020,He2021}.

The binding energy where valence and conduction band states originating from the different AB layers in tDBG hybridise can be shifted from the Fermi energy by imposing large twist angles.  In this case,  application of an out-of-plane electric field via top and bottom dual gate electrodes leads to the possibility of separately charging the AB layers with electrons and holes,  thereby building up a large interlayer polarization \cite{Rickhaus2019}.  If the twist angle is gradually reduced below 4$^{\circ}$,  hybridisation between the electronic bands occurs closer to the Fermi energy,  which flattens the dispersion and enhances the correlation strength \cite{Jones2020,Folkert:2020,Tomic:2022}.  There is a cross-over regime in tDBG where enhanced correlations combined with significant interlayer polarization, induced via gating, can lead to strong nesting of electron- and hole-type bands in the separate AB layers, potentially inducing charge density wave or electron-hole (exciton) insulating states \cite{Rickhaus:2021}.  The electronic structure of tDBG in this cross-over regime of displacement fields and twist angles has not been studied,  leaving open questions about the size and tunability of hybridisation gaps and flat bands that constitute the necessary ingredients for achieving such states.  We address these questions here using micro-focused angle-resolved photoemission spectroscopy (microARPES) on \textit{in situ} gated tDBG devices with moderate twist angles that fall into the cross-over regime from weak-to-strong correlations.  Our observations underscore the potential of tDBG for realizing strongly correlated electron-hole states that could lead to exciton condensation as well as superconductivity driven by excitons \cite{Zefang:2019,Su2023}.

\section{Results and Discussion}

\begin{figure*} [t!]
	\begin{center}
		\includegraphics[width=1.0\textwidth]{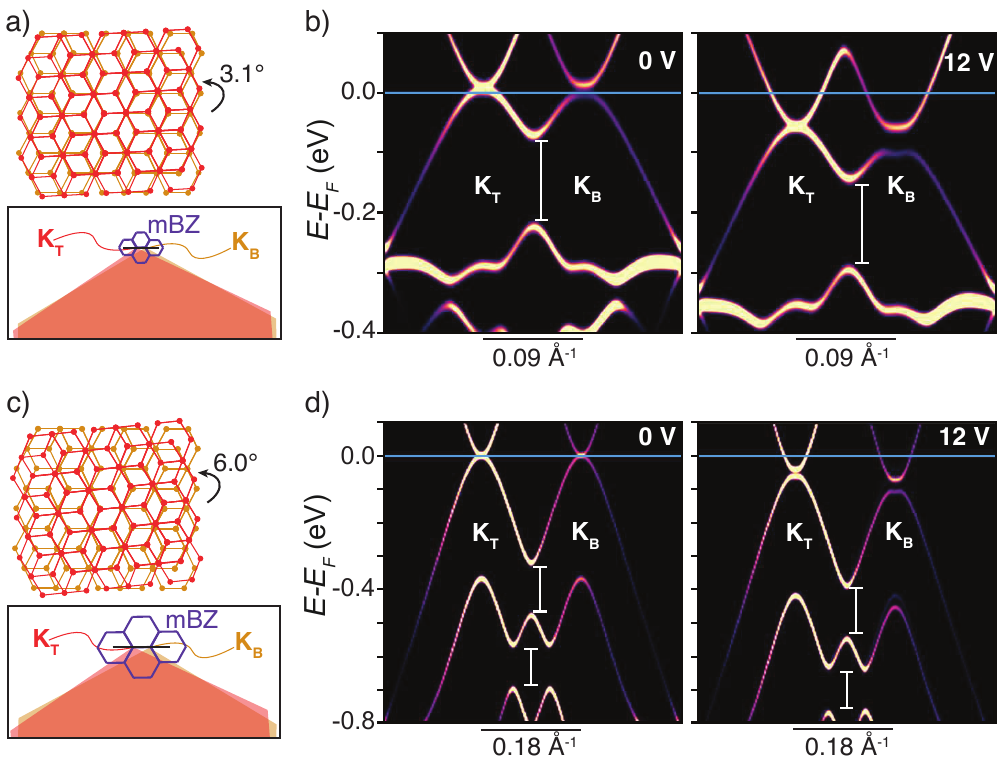}
		\caption{Tuning the electronic structure of tDBG with twist angle and electrostatic gating.  a) Structure and BZs of two AB bilayers of graphene with an interlayer twist angle of 3.1$^{\circ}$.  Four mBZs of the resulting superstructure are shown. The top (bottom) AB bilayer is shown in red (brown).  b) Calculated spectral functions of 3.1$^{\circ}$ tDBG in the presence of a bottom gate voltage of 0 and 12~V along the cut indicated by a black line in the mBZs in (a).  c) Structure,  BZs and mBZs for 6.0$^{\circ}$ tDBG.  d) Calculated spectral functions of 6.0$^{\circ}$ tDBG at the stated gate voltages.  The white brackets in (b) and (d) demarcate hybridization gaps.}		
	\label{Fig:1}
	\end{center}
\end{figure*}

The material systems and electronic structures that form the basis of our study are introduced in Fig.  \ref{Fig:1}.  Two AB graphene bilayers are exfoliated and stacked with an interlayer twist angle, which is set to a minimum of 3.1$^{\circ}$ in this study (see Fig.  \ref{Fig:1}(a))  as this corresponds to the regime where correlation effects and moiré physics starts to impact the transport properties of tDBG \cite{Rickhaus:2021}.  Twisting the AB layers corresponds to adding a rotation angle between the Brillouin zone (BZ) of each layer such that top and bottom $\mathrm{K}$-points,  denoted as $\mathrm{K_T}$ and $\mathrm{K_B}$, are separated by 0.09~\AA$^{-1}$.  The BZs of each layer and the resulting mini Brillouin zone (mBZ) are sketched in Fig.  \ref{Fig:1}(a) along with a black bar that demarcates the $\mathrm{K_T}$-$\mathrm{K_B}$ direction, which straddles the four mBZs in the sketch.  

Calculated spectral functions unfolded along the $\mathrm{K_T}$-$\mathrm{K_B}$ cut for the same twist angle, dielectric conditions and gate voltages that we realize in our experiment are shown in Fig.  \ref{Fig:1}(b),  revealing two separate bands at the Fermi level $E_\mathrm{F}$ and a large hybridisation gap between them (see brackets in Fig.  \ref{Fig:1}(b)).  As the gate voltage is applied the bands in the lower layer shift down,  the band gap increases and the valence band maximum (VBM) develops a subtle M-shape.  Simultaneously,  the conduction band minimum (CBM) in the lower layer aligns in energy with the gapless VBM-CBM region of the top layer.  A difference in the charge neutrality point (CNP) energy between the layers is possible because of the inherent asymmetry of having the bottom bilayer supported on hBN and top layer facing vacuum.  This sets up an interlayer charge polarization and thus an electric displacement field in response to the applied electric field.  

Creating a tDBG sample with an interlayer twist angle of 6.0$^{\circ}$ leads to a larger mBZ (see Fig.   \ref{Fig:1}(c)) with a  $\mathrm{K_T}$-$\mathrm{K_B}$ separation of 0.18 \AA$^{-1}$.  In this scenario,  the calculations reveal bands that hybridise at substantially lower energies than observed for 3.1$^{\circ}$ tDBG,  around -0.4 eV (see Fig.   \ref{Fig:1}(d)).  The resulting hybridisation gaps and hybridised segments of bands merely exhibit subtle shifts to lower energies with applied gate voltages.

\begin{figure*} [t!]
	\begin{center}
		\includegraphics[width=1.0\textwidth]{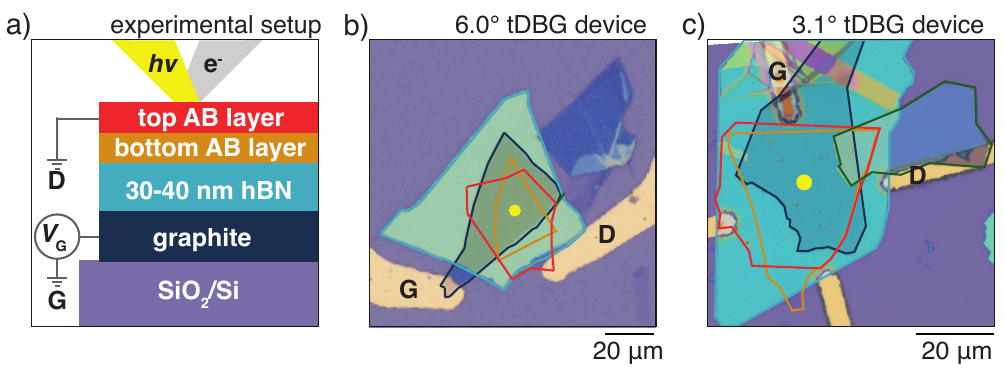}
		\caption{Experimental setup and devices.  a) Schematic of photoemission experiment and heterostructure composition.  b)-c) Optical images of the (b) 6.0$^{\circ}$ and (c) 3.1$^{\circ}$  tDBG devices.  The top and bottom AB layers are indicated by red and brown outlines, respectively, while the graphite and hBN flakes are marked by dark and light blue outlines.  In (c),  a spare graphite layer picked up together with the back-gate is outlined in green. The yellow circle indicates the areas of the sample that were measured in the micro-focused photoemission experiments.  Gate and drain electrodes are annotated as "G" and "D".}
		\label{Fig:2}
	\end{center}
\end{figure*}

Figure \ref{Fig:2}(a) presents our microARPES experimental setup and the composition of our tDBG heterostructure devices \cite{Jones:2025}.  Two twisted AB graphene layers are deposited on hBN with a thickness of 30-40 nm.  The stack is placed on a graphite back gate on SiO$_2$/Si.  The top AB layer is connected to ground via a drain electrode.  Photoemission is performed at the ASTRID2 light source at Aarhus University, Denmark, using photon energy $hv = 47$~eV.  All devices are kept at a temperature of 300 K during the microARPES measurements.

Our 6.0$^{\circ}$ and 3.1$^{\circ}$ tDBG devices are presented via optical images in Figs.  \ref{Fig:2}(b)-(c): Top and bottom AB flakes are demarcated by red and brown lines, respectively, while the graphite and hBN flakes are delineated in dark and light blue.  The bilayer graphene flakes are not aligned with the hBN. The graphite flake used as gate electrode has been positioned directly under the part of the tDBG stack that is exposed to the microscopic light spot (see yellow circle in  Figs.  \ref{Fig:2}(b)-(c)) during microARPES measurements.   The overall configuration of graphite electrode,  and gate/drain leads has been designed to minimize inhomogeneous stray electric fields that distort the measured kinetic energy and momentum distributions of photoemitted electrons.

\begin{figure*}[t!] 
\begin{center}
\includegraphics[width=1.0\textwidth]{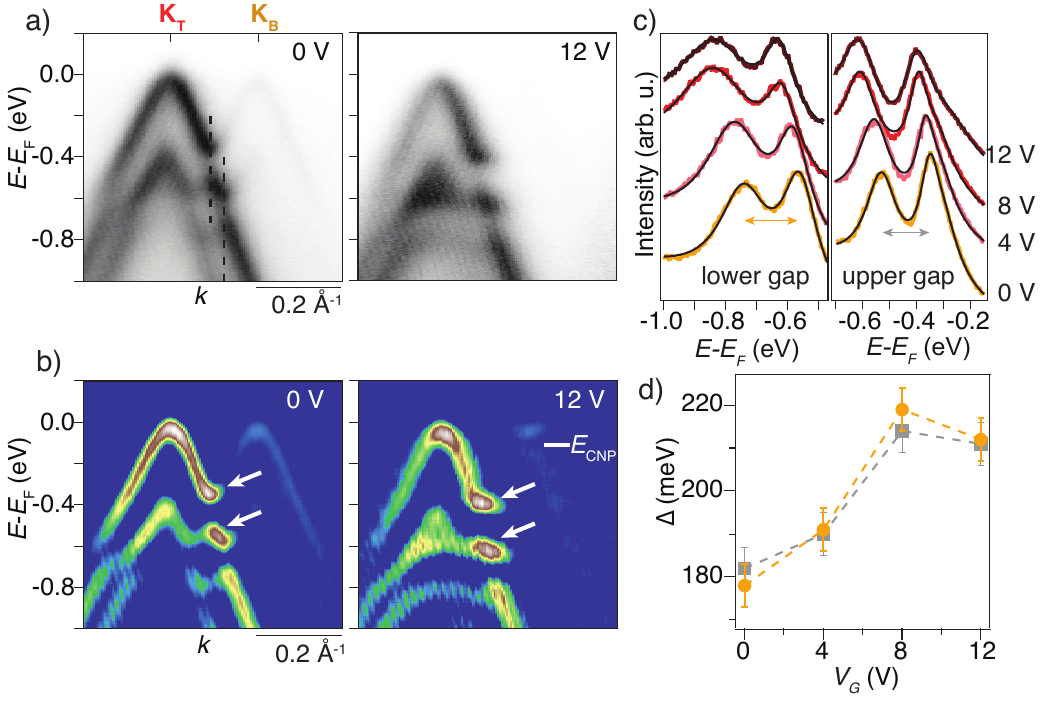}
\caption{Hybridisation and electric-field dependent gaps in 6.0$^{\circ}$ tDBG.  a) ARPES dispersion along the $\mathrm{K_T}-\mathrm{K_B}$ cut at 0 and 12~V. b) Second derivative ARPES intensity corresponding to the spectra in (a).  The charge neutrality point $E_{\mathrm{CNP}}$ in the bottom AB layer is demarcated by a white tick.  The white arrows demarcate band segments that change slope with increasing gate voltage.  c) EDCs (colored curves) extracted along the dashed vertical lines shown on the 0 V spectrum in (a),  representing measurements at 0, 4, 8 and 12~V.  The double-headed arrows demarcate the lower and upper hybridisation gaps.  Fits to two Lorentzian peaks on a linear background are overlaid (black curves).  Peak-to-peak energy difference $\mathrm{\Delta}$ corresponding to the upper (grey squares and dashed lines) and lower (orange circles and dashed lines) hybridisation gaps as a function of gate voltage. }
\label{Fig:3}
\end{center}
\end{figure*}

To directly visualize how an out-of-plane electric field modifies interlayer hybridization in tDBG, we first compare ARPES dispersions of the 6.0$^{\circ}$ device along the $\mathrm{K_T}-\mathrm{K_B}$ cut at different gate voltages as shown in Fig.  \ref{Fig:3}(a). This allows us to identify the two band crossings and the corresponding upper and lower hybridization gaps.  For each bilayer in the heterostructure, we observe two sets of bands separated in energy by 0.4 eV around the $\mathrm{K}$-points. The energy separation is in line with ARPES measurements on pristine AB bilayer graphene \cite{Ohta:2006}.  The $\mathrm{K}$-points are separated by 0.18~\AA$^{-1}$,  as sketched in Fig.  \ref{Fig:1}(c). The ARPES intensity emerging from the top AB layer is 14 times higher than in the lower AB layer due to the short inelastic mean free path of photoexcited electrons with kinetic energy around 40 eV.  The upper bands intersect and form an upper hybridisation gap centred at an energy of -0.4~eV, whereas the lower bands intersect and form a lower hybridisation gap centred at an energy around -0.7~eV (see vertical dashed lines in Fig.   \ref{Fig:3}(a) at 0~V).  Upon increasing the gate voltage,  we measure ARPES spectra at 4~V,  8~V (not shown) and 12 V.  We could not achieve higher gate voltages without observing detrimental distortions in the ARPES spectra, caused by stray electric fields.  

The spectrum measured at 12 V is presented next to the 0~V data in Fig.   \ref{Fig:3}(a).  Because the bottom layer bands have much weaker ARPES intensity, we use second derivative images to enhance faint dispersive features and to highlight subtle changes in band flattening near the hybridization regions as shown in Fig.  \ref{Fig:3}(b).  We observe the CNP energy of the bottom AB layer has shifted below the Fermi energy,  as indicated by a white tick in Fig.  \ref{Fig:3}(b).   The top AB layer shifts very little,  which indicates that most of the field-induced charge is screened by the bottom AB layer.  The band segments that intersect change slope and significantly flatten as seen via white arrows in Fig.  \ref{Fig:3}(b).  It is thus possible to realize a voltage tunable increase in the density of states (DOS) around energies of -0.4 and -0.7 eV.  

To quantify how the hybridization gap evolves with gate-induced displacement field,  we extract energy distribution curves (EDCs) at fixed momenta near the band intersections (see vertical dashed lines in Fig.  \ref{Fig:3}(a)) and track the peak-to-peak separation $\Delta$ as a function of gate voltage.  EDCs with fits to two Lorentzian peaks are shown in Fig.  \ref{Fig:3}(c) with double-headed arrows demarcating the gap sizes $\mathrm{\Delta}$.  These are plotted versus gate voltage in  Fig.  \ref{Fig:3}(d),  which shows that an increasing electric field (and thus displacement field) drives an increase in $\mathrm{\Delta}$ from around 180 to 210 meV.  Both gaps are similar in magnitude for a given applied field.  The increase in $\mathrm{\Delta}$ may be explained by the observed band flattening effects where the bands hybridise.  The reduced slope of the bands concentrates the centroid of the intensity further away from the middle of the gap.  While the general features of the band structure and doping-induced shift of the dispersion are in line with the spectral function calculations in Fig.  \ref{Fig:1}(d),  the subtle changes of band slope that drives band flattening and increases $\mathrm{\Delta}$ are not captured,  as such effects are likely highly sensitive to the interlayer tunnelling strength.  We speculate that such flat segments could lead to sharper optical transitions between the conduction and valence bands.  Assuming that these features develop at similar energies in the unoccupied states, such band segments with enhanced DOS would be resonant with infrared optical transitions and potentially contribute to non-linear components in the optical conductivity as well as the generation of excitons in tDBG \cite{Havener:2014,Liang:2014aa}.

\begin{figure*}[t!] 
\begin{center}
\includegraphics[width=1.0\textwidth]{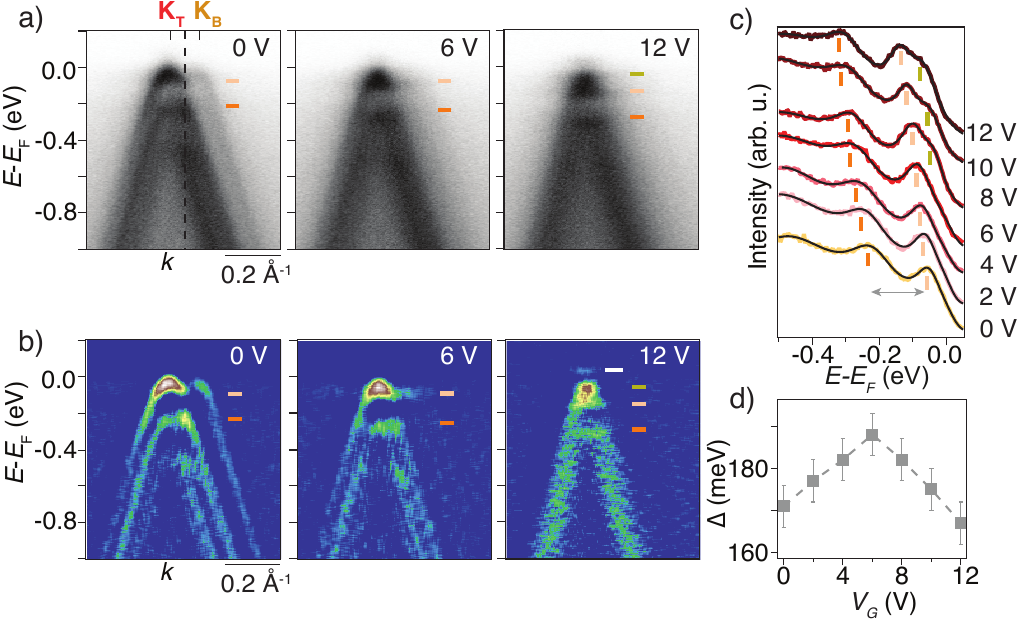}
\caption{Electric field-driven dispersion changes in 3.1$^{\circ}$ tDBG.  a)-b) ARPES dispersion of hyperbolic TDBG bands along $\mathrm{K_T}-\mathrm{K_B}$ at 0,  6 and 12~V presented via the (a) raw ARPES intensity and (b) second derivative of the ARPES intensity.  An enhanced state in the CB region of the top AB layer is demarcated by a white tick at 12~V in (b).  c) EDCs (colored curves) measured at the given values of gate voltage along the vertical dashed line shown on the 0 V spectrum in (a).  Fits to two or three Lorentzian peaks on a linear background are overlaid (black curves).  The ticks in (a)-(c) demarcate the peak positions of the fitted Lorentzians,  corresponding to flat segments on the dispersion.  (d) Hybridisation gap as a function of applied gate voltage.  The double-headed arrow in (c) demarcates the peak-to-peak energy separation $\mathrm{\Delta}$ that is used as a measure of the gap size. }
\label{Fig:4}
\end{center}
\end{figure*}

Figure \ref{Fig:4} presents ARPES spectra of 3.1$^{\circ}$ tDBG at 0,  6 and 12 V. The device configuration is very similar to the 6.0$^{\circ}$ tDBG device,  such that the same limitations in maximum achievable gate voltage are experienced. The two sets of bands are separated in $k$ by 0.09 \AA$^{-1}$,  giving rise to a small mBZ as sketched in Fig. \ref{Fig:1}(a). Hybridisation occurs below the VBM,  around -0.1 eV,  forming a quasi-flat band separated by a hybridisation gap from another flat band segment below -0.2 eV.  With increasing gate voltage the lower AB layer band structure exhibits a reduced ARPES intensity and its dispersion becomes difficult to discern, but interestingly, the intensity is enhanced at energies where hybridisation occurs with the top layer bands.  Further details may be discerned from second derivative images of the ARPES measurements as shown in Fig.  \ref{Fig:4}(b).   At 12 V,  the CBM of the top layer is visible as a faint flat band just above the Fermi level (see white tick in Fig.  \ref{Fig:4}(b)).  It is observable, albeit with low intensity,  because the measurements have been performed at 300 K,  leading to a significant thermal population of states above $E_\mathrm{F}$.  This direct observation of the top layer CBM provides an important clue to the origin of the displacement field-driven changes in the lower-lying manifold of bands.  In particular,  the band segments below $E_\mathrm{F}$ originating from the top AB bilayer must stem from hybridization-induced changes of the VBM with the bands of the bottom AB bilayer. 

We interpret the observed dispersion changes around the CBM and VBM as the result of flat band formation induced by the applied gate voltage.  The flat band segments are highlighted by ticks in Figs.  \ref{Fig:4}(a)-(b).  In order to locate the energy of the flat bands and determine the hybridization gaps separating them,  we extract EDCs along the vertical dashed line in Fig.  \ref{Fig:4}(a) as a function of gate voltage.  Data with fits composed of Lorentzian peaks are presented in Fig.  \ref{Fig:4}(c).  Two peaks straddle a hybridisation gap up to 6 V.  At 8 V, a third peak develops closer to $E_\mathrm{F}$, which shifts to lower energies as the gate voltage is further increased.  The peak-to-peak gap energy extracted from the EDCs is shown in Fig.  \ref{Fig:4}(d).  The gap is seen to increase up to 6 V,  but it then decreases as the field is further increased.  Thus,  with increasing applied field a non-monotonic behaviour of the hybridization gap is observed.

\begin{figure*}[t!] 
\begin{center}
\includegraphics[width=1\textwidth]{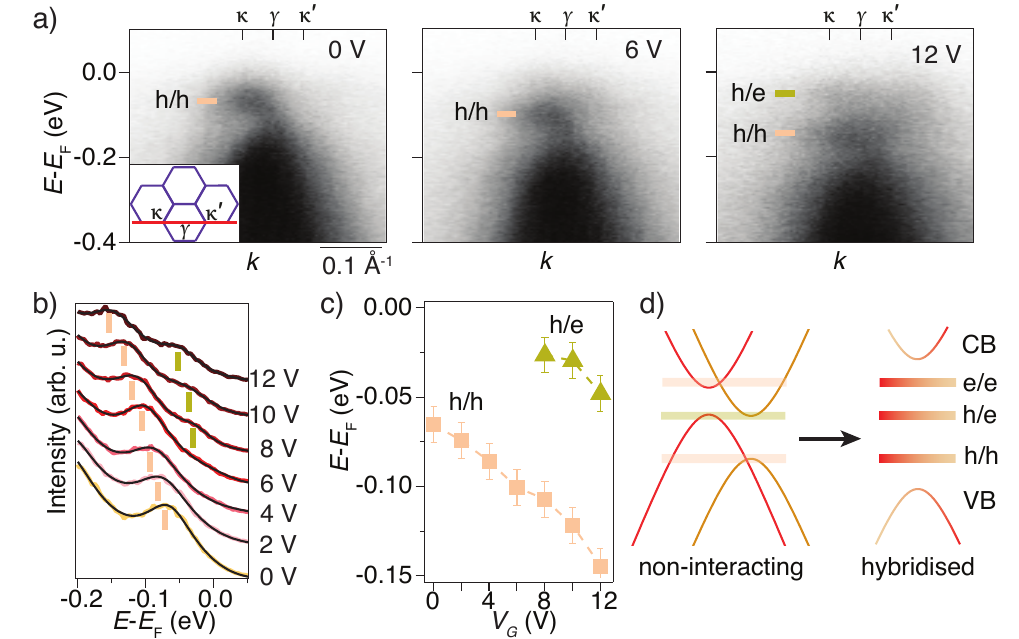}
\caption{Mixed electron- and hole-like flat bands in 3.1$^{\circ}$ tDBG.  a) ARPES spectra at the stated gate voltages along the $\kappa-\gamma-\kappa^{\prime}$ direction demarcated by a red line in the inset with four mBZs.  b) EDCs (colored curves) binned between  $\kappa$ and $\gamma$, corresponding to a $k$-range of 0.05~\AA$^{-1}$,  at the given values of gate voltage.  Fits to one or two Lorentzian peaks on a linear background multiplied by a Fermi-Dirac function are overlaid (black curves).  Fitted Lorentzian peak positions are demarcated by ticks in (a)-(b).  c) Fitted peak positions as a function of gate voltage.  d) Diagrams of non-interacting low-energy hyperbolic bands of top (red) and bottom (brown) AB layers along with the resulting hybridised band structure.  The overlaid bars on the non-interacting bands indicate the segments of the band structures that hybridise.  Mixed electron- and hole-like hybridised bands are denoted h/e and pure hole- or electron-like hybridised bands are denoted h/h and e/e,  respectively.}
\label{Fig:5}
\end{center}
\end{figure*}

The impact of the  electric field is investigated in further detail in Fig.  \ref{Fig:5}(a),  which presents a cut of the ARPES intensity along the $\kappa-\gamma-\kappa^{\prime}$ direction of the mBZ (see red bar in inset with mBZs in Fig. \ref{Fig:5}(a)).  We observe that the hybridised part of the upper and lower layer valence bands form a flat band along this part of the mBZ.  Upon increasing the gate voltage the flat band shifts down and eventually an additional flat band can be observed closer to $E_\mathrm{F}$.  The energies of the flat bands are determined by fitting Lorentzian peaks to EDCs binned across $\kappa-\gamma-\kappa^{\prime}$,  as shown in Fig.  \ref{Fig:5}(b).  Fitted flat band positions are shown as a function of gate voltage in Fig.  \ref{Fig:5}(c).  The flat bands originate from a combination of electric field induced dispersion changes in each layer and the downwards shifts of the lower layer bands, opening new possibilities for hybridisation across the two layers.  Figure \ref{Fig:5}(d) presents a schematic of the situation.  As shown in the calculated band structure in Fig.  \ref{Fig:1}(b) the CNP of the lower layer shifts significantly below the upper layer.  This behavior is directly evident in the 6.0$^{\circ}$ tDBG sample at 12 V in Fig.  \ref{Fig:3}(b).  In the 3.1$^{\circ}$ tDBG device this shift of the CNP entails hybridization between the VBM of top AB bilayer and the CBM of the bottom AB bilayer. Consequently,  as the CBM of the bottom AB layer shifts below the energy of the VBM of the top layer,  these band segments hybridise and form a mixed electron- and hole-like flat band, which we denote as h/e.  At lower energies the set of valence bands form a hole-hole like flat band (denoted h/h),  whereas the emergence of the CB state at 12 V (see Fig. \ref{Fig:4}(b)) indicates the formation of an electron-electron like flat band (denoted e/e) just above $E_{\mathrm{F}}$.  The flattening of the dispersion with gate voltage is not captured in the spectral function calculation in Fig.  \ref{Fig:1}(b).  Similarly as for the 6.0$^{\circ}$ sample where slope changes were not captured,  the details of the dispersion changes in the 3.1$^{\circ}$ structure are connected with strong interlayer tunnelling and significant changes of the on-site potential driven by increasing displacement field. 

The experimentally observed flat band segments indicate an enhanced DOS around $E_{\mathrm{F}}$,  which poses 3.1$^{\circ}$ tDBG as an interesting material to explore in low-temperature transport experiments to uncover correlation effects involving flat bands with mixed electron- and hole-type character.  While our ARPES experiment does not resolve the detailed dispersion within the h/e band,  we speculate that hole- and electron-like bands may overlap in energy.  Analogous to the situation in the semimetals TiSe$_2$ and Ta$_2$NiSe$_5$ \cite{Cercellier:2007,Lu:2017},  this scenario could lead to the spontaneous formation of excitons,  thereby forming an unconventional insulating state in the form of an excitonic insulator \cite{Rickhaus:2021}.  Another scenario that we may speculate involves band inversion in the h/e state and the formation of non-trivial topological states with Chern number exchange between mini-valleys that give rise to anomalous Hall effect signatures in transport experiments \cite{PhysRevB.99.235406,PhysRevB.99.075127,Lee2019,Shaowen:2021}.  Our observation of a h/e flat band in 3.1$^{\circ}$ tDBG thus opens the possibilty of realising moiré semimetals with correlated and topological states at moderate twist angles, which soften the experimental constraints associated with achieving magic angle devices.

\section{Conclusion}
We applied microARPES on in situ gated $6.0^{\circ}$ and $3.1^{\circ}$ tDBG in order to investigate the impact of a vertical electric field on the electronic structure across twist angles where interlayer polarization and band flattening impact the optical and electronic properties.  In $6.0^{\circ}$ tDBG,  we find that several hybridisation gaps form 0.4 eV below the Fermi energy between the  bands from each AB layer.  The hybridised band segments flatten and the gap sizes increase as a larger electric field is induced.  In the case of $3.1^{\circ}$ tDBG,  we observe band flattening and a non-monotonic hybridisation gap with increasing external field.  Flat bands develop between valence and conduction band extrema in the two layers,  thereby mixing electron- and hole-type carriers in narrow bands that are accessible in low-energy transport experiments.

\section{Experimental Section}

\textit{Device fabrication} \\
Bilayer graphene, hBN and graphite flakes were exfoliated onto SiO$_2$/Si substrates and identified by optical microscopy. The heterostructures were assembled using a conventional dry-transfer technique \cite{Kyounghwan:2016, Kyounghwan:2017} based on a polycarbonate (PC)/polydimethylsiloxane (PDMS) stamp mounted on a glass slide. Two bilayer graphene flakes were each cut into two pieces using a tungsten micro-needle.

One piece from each bilayer graphene flake was picked up at 130~$^{\circ}$C by the polymer stamp. The remaining piece was then rotated by a defined angle and subsequently picked up using the first piece, forming a twisted double bilayer. For the first device the twist angle was $3.1^{\circ}$, and for the second device it was $6.0^{\circ}$. After assembling the twisted graphene, an hBN flake was picked up at 110~$^{\circ}$C, followed by a graphite flake at 130~$^{\circ}$C. The complete stack was finally released onto a SiO$_2$/Si substrate pre-patterned with Au/Cr electrodes (50/5~nm) at 180~$^{\circ}$C. The devices were annealed in a H$_2$/Ar atmosphere at 250~$^{\circ}$C for 4~hours and subsequently wire-bonded to a chip carrier.\\

\textit{Photoemission measurements} \\
MicroARPES measurements were carried out at the SGM4 beamline of the ASTRID2 synchrotron radiation source at Aarhus University, Denmark \cite{Volckaert:2023, Jones:2025}. The devices were wire-bonded to CSB00815 chip carriers and annealed in the microARPES end-station at 180~$^{\circ}$C for 1~hour prior to exposure to the beam. During data acquisition, the samples were kept at room temperature in a base pressure better than $2 \times 10^{-10}$~mbar.

A synchrotron beam with a photon energy of 47~eV was focused to a spot size of approximately 2.5~$\mu$m using an elliptical capillary mirror. The ARPES spectra were recorded with a SPECS Phoibos 150 SAL analyser. The energy and angular resolutions were better than 20~meV and 0.1$^{\circ}$, respectively. $(E, k_x, k_y)$-dependent photoemission intensity maps were acquired using the scanning angle lens mode of the analyser while keeping the sample position fixed. Four-dimensional $(E, k, x, y)$ datasets were obtained by rastering the sample position relative to the beam with a piezoelectric stage  and recording an $(E, k)$ snapshot at each spatial point. Electrostatic doping of the devices \textit{in situ} was controlled using Keithley 2450 source meters.\\

\textit{Calculations:}\\
Commensurate twisted double bilayer graphene (tDBG) structures with twist angles of \(3.1^{\circ}\) and \(6.0^{\circ}\) were relaxed using classical molecular dynamics in \textsc{LAMMPS}~\cite{thompson_lammps_2022} with the REBO2~\cite{brenner_second-generation_2002} and EXX-RPA-fitted DRIP potentials~\cite{leconte_relaxation_2022}. Each model consisted of two AB-stacked bilayer graphene layers (lattice constant \(a = 2.46~\text{\AA}\)). The relaxed structures were used to construct an extended Hubbard Hamiltonian:  
\[
H = H^0 + \sum_{i\sigma}(U\Delta\rho_{ii\tilde\sigma} + \sum_j V_{ij}\Delta\rho_{jj})c^{\dagger}_{i\sigma}c_{i\sigma} - \sum_{ij\sigma}V_{ij}\Delta\rho_{ij\sigma}c^{\dagger}_{i\sigma}c_{j\sigma},
\]
where \(H^0 = \sum_{ij\sigma} t_{ij} c^{\dagger}_{i\sigma} c_{j\sigma}\). The difference matrix \(\Delta\rho\) removes double-counting of Coulomb effects.

Hopping models \(t_{ij}\) were constructed using the scaled hybrid exponential model~\cite{leconte_relaxation_2022} and extended Hubbard corrected hopping parameters~\cite{lee_extended_2024}. The on-site interaction was fixed to \(U = 6.2~\text{eV}\); intersite terms \(V_1 = 3.2~\text{eV}\) and \(V_2 = 2.1~\text{eV}\) were used for first and second intralayer neighbors, and interlayer interactions followed  
\[
V_{ij} = 1/(\epsilon_r r_{ij}),
\]
with \(\epsilon_r = 2.7\). The cutoff for the interlayer interaction is set to \(4.9~\text{\AA}\). All self-consistent calculations employed a \(13 \times 13\) \(k\)-point mesh.

The unfolded spectral function was computed using the band-unfolding formalism of Ref.~\cite{nishi_band-unfolding_2017}:  
\[
A(k,E) = -\frac{1}{\pi}\frac{S_g}{S_m}\sum_{K,G,N}\delta_{k-G,K}\sum_s \omega^s_{KN}(k)\operatorname{Im}\!\left[\frac{1}{E-\epsilon_{KN}+i\eta}\right],
\]
where \(S_g\) and \(S_m\) are the graphene and moiré unit-cell areas. The sublattice-resolved spectral weight is given by  
\[
\omega^s_{KN}(k) = \sum_{I,J \in s}\delta_{i(I),i'(J)} (C^{KN}_I)^*C^{KN}_J e^{ik(T_s(I) - T_s(J))},
\]
where \(I\) and \(J\) label orbital indices, \(C^{KN}_I\) are eigenvector components, and \(T_s(I)\) are lattice translation vectors of sublattice \(s\).

\section{acknowledgement}
The work was funded/co-funded by the European Union (ERC grant EXCITE with project number 101124619). Views and opinions expressed are
however those of the author(s) only and do not necessarily reflect those of the European Union or the European
Research Council.  Neither the European Union nor the granting authority can be held responsible for them.  The authors acknowledge funding from the Novo Nordisk Foundation (Project Grant NNF22OC0079960),  VILLUM FONDEN under the Villum Investigator Program (Grant. No. 25931),  and the Danish Council for Independent Research (Grant Nos.  DFF-9064-00057B and 1026-00089B).  D.L.  acknowledges support by the National Research Foundation of Korea (NRF), RS-2024-00413481.  J.J.  and Y.P.  acknowledge support from the National Research Foundation of Korea (NRF), NRF-2020R1A5A1016518. C.S. acknowledges the Marie Sklodowska-Curie Postdoctoral Fellowship (proposal number 101059528).  K.W. and T.T. acknowledge support from the JSPS KAKENHI (Grant Nos. 21H05233 and 23H02052) and World Premier International Research Center Initiative (WPI), MEXT, Japan.  We acknowledge the Urban Big Data and AI Institute of the University of Seoul supercomputing resources (http://ubai.uos.ac.kr).

\section{Data Availability}
The data used in this study is available on Zenodo \cite{Zenodo}.

\section{Keywords}
Twisted double bilayer graphene,  moir\'e superlattice, heterostructures, microARPES, electronic structure,  electric field-driven flat bands

\end{document}